\documentclass[]{pasj02} 
\usepackage[switch,mathlines]{lineno} 
\usepackage{bm}  
\Received{2026/1/13}
\Accepted{2026/3/3}

\begin{document} 

\title{Hub Formation and Filament-Filament Collision: An Analytical Model}

\author{
Kohji \textsc{Tomisaka},\altaffilmark{1,2}\altemailmark\orcid{0000-0003-2726-0892} \email{kohji\_tomisaka@yahoo.co.jp} 
Raiga \textsc{Kashiwagi},\altaffilmark{1,3}\orcid{0000-0002-1461-3866}
{ }and
Kazunari \textsc{Iwasaki}\altaffilmark{1,2} \orcid{0000-0002-2707-7548}
}
\altaffiltext{1}{National Astronomical Observatory of Japan,
Osawa 2-21-1, Mitaka, Tokyo 181-8588, Japan}
\altaffiltext{2}{Astronomical Science Program, Graduate University for Advanced Studies,
Osawa 2-21-1, Mitaka, Tokyo 181-8588, Japan}
\altaffiltext{3}{Korea Astronomy and Space Science Institute, 
Daejeon 34055, Republic of Korea}




\maketitle

\begin{abstract}
Filaments are ubiquitous throughout the Galaxy.
Massive star formation is often observed in hub-filament systems, where multiple filaments appear to be interconnected and merging. 
Filament-filament collisions are therefore a likely triggering mechanism for massive star formation.
We derive basic physical properties of filament-filament collisions, 
 such as the collision cross section (CCS), the hub mass, and its mass function, based on a simple cylindrical filament model. 
We assume a cylindrical filament with length $2p$, full width $2q$, and line-mass $\lambda_0$,
 and consider the CCS between two identical filaments.
The collision is specified by three vectors: 
 the directions of the colliding filaments ($\bm{n}_1$ and $\bm{n}_2$) and the direction of the relative velocity between the two filaments 
 ($\bm{n}_v=\bm{v}/|\bm{v}|$).
For the thin filament, $p\gg q$, 
 the CCS is expressed as $S=4p^2|\bm{n}'_1\times \bm{n}'_2|$,
 where $\bm{n}'_1$ and $\bm{n}'_2$ represent the directional vectors projected onto a plane perpendicular to the relative velocity $\bm{n}_v$. 
As the angle 
between $\bm{n}'_1$ and $\bm{n}'_2$ becomes smaller, 
 the cross section 
 proportional to $p\cdot q$ becomes relatively important. 
We propose a simple model in which the hub mass is 
 estimated by 
 the overlapping portion of the two colliding filaments.
The hub mass function is derived using the CCSs and the geometrically estimated overlapping mass.
When the directions and relative velocities of the filaments are isotropically distributed,
the mass function expected from a single species of filaments fits well to a power law
and the power exponent is $\gamma_M\simeq -2.96$ -- $-3.78$.
The power exponent of the global hub mass function is the same as 
 that of the line-mass distribution function, 
$\gamma_\lambda\simeq -1.5$.
This means that a massive hub is formed by the collision of two massive filaments. 
\end{abstract}


\section{Introduction} \label{sec:intro}
Filaments have been attracting continuous attention since the Herschel satellite found that molecular clouds consist
 of filamentary structures  (for a review, see \authorcite{2014prpl.conf...27A} \yearcite{2014prpl.conf...27A}; 
 \authorcite{2023ASPC..534..153H} \yearcite{2023ASPC..534..153H}; 
 \authorcite{2023ASPC..534..233P} \yearcite{2023ASPC..534..233P}). 
In less massive star-forming regions, protostars and bound starless cores often delineate the filaments found 
 in the far IR thermal emissions from dust grains \citep{2010A&A...518L.102A}. 
On the other hand, the formation of massive stars is mainly seen in high-density hubs where multiple filaments intersect each other 
\citep{2009ApJ...700.1609M,2020A&A...642A..87K}, which are called hub-filament structures or systems (HFSs). 
This leads to the idea that the collision or coalescence of multiple filaments may initiate the formation of massive stars
 by forming high-density gaseous hubs (e.g \cite{2014ApJ...791L..23N}; \cite{2019ApJ...875..138D};
 \cite{2022A&A...660L...4B};
 \cite{2022MNRAS.513.2942D};
 \cite{2022A&A...658A.114K};
 \cite{2022ApJ...934....2M};
 \cite{2022MNRAS.515.1012Y};
 \cite{2023MNRAS.519.2391Z};
 \cite{2024MNRAS.527.4244S}; 
 \cite{2024MNRAS.527.5895D}; 
 \cite{2024A&A...686A.146Z}; 
 \cite{2024ApJ...967..151S};  
 \cite{2024A&A...688A..99Z}; 
 \cite{2025AJ....169...56M}; 
 \cite{2025AJ....169...80D}; 
 \cite{2025ApJ...986...48J}). 

To form massive stars before reaching their main-sequence stage requires a much larger mass accretion rate such as $\dot{M}_*\simeq 10^{-4}-10^{-3}\,M_\odot \mathrm{yr}^{-1}$ than that for less massive stars (e.g. \cite{2003ApJ...585..850M}).
\citet{2003ApJ...585..850M} consider a massive clump supported mainly by turbulent pressure and confined with extremely high ambient pressure and show that a high accretion rate is expected in such a dense turbulent clump.
The essence is to form a clump with a mass $M_\mathrm{turb}$ much greater than the thermal Jeans mass $M_\mathrm{J}$.
The accretion rate is given by the mass $M_\mathrm{turb}$ and the free-fall time of the gas $t_\mathrm{ff}$ as $\dot{M}_*\sim M_\mathrm{turb}/t_\mathrm{ff} \gg M_\mathrm{J}/t_\mathrm{ff} = \rho c_s^3 t_\mathrm{ff}^2 \sim c_s^3/G$, where $c_s$, $\rho$, and $G$ represent respectively the isothermal sound speed of the interstellar gas with a temperature, $T\sim 10\,\mathrm{K}$, gas density, and the gravitational constant, and $c_s^3/G\sim 10^{-6} (T/10\,\mathrm{K})^{3/2}\,M_\odot \mathrm{yr}^{-1}$ is approximately equal to the mass accretion rate expected for the formation of less massive stars \citep{1977ApJ...214..488S,1985MNRAS.214....1W,1996PASJ...48L..97T}.  
By filament-filament collision, a dense hub is likely to be formed in a relatively short crossing time scale determined
 by the width of the filament $D$ and the collision speed $V$, as
 $\tau \sim D/V\simeq 3\times 10^5 \mathrm{yr} (D/ 0.3\,\mathrm{pc})(V/1 \mathrm{km\,s^{-1}})^{-1}$.
This results in a prompt mass collection larger than the thermal Jeans mass.  

The hub mass predicted by a filament-filament collision with line-mass of $\lambda_0$ is estimated as 
 $M_\mathrm{hub}\gtrsim 2\lambda_0 D \simeq 60 (\lambda_0/100\,M_\odot\,\mathrm{pc}^{-1}) (D/0.3\,\mathrm{pc})\,M_\odot$
 (Although the mass depends on the collision direction, this value represents the minimum).
 The mass collection rate reaches
 $\dot{M}_c\sim M_\mathrm{hub}/\tau
 \sim 200 (\lambda_0/100\,M_\odot\,\mathrm{pc}^{-1}) (V/1 \,\mathrm{km\,s^{-1}})\,M_\odot\,\mathrm{Myr}^{-1}$,
 which is comparable to the mass inflow rate observed along the filament axis in HFSs
 (e.g. \cite{2013ApJ...766..115K,2019ApJ...875...24C,2023A&A...676A..15M,2024MNRAS.528.2199R}).     

A number of (magneto-)hydrodynamical simulations have been performed regarding the filament-filament collisions \citep{2011A&A...528A..50D,2021MNRAS.507.3486H,
2023ApJ...954..129K,2024ApJ...974..265K,10.1093/mnrasl/slae045}.  
The outcome of the collision between two parallel filaments is determined
 by whether the total line-mass exceeds the critical line-mass or not \citep{2021MNRAS.507.3486H,2023ApJ...954..129K}\footnote{For the critical line-mass of cylindrical filaments without a magnetic field, see \citet{1963AcA....13...30S,1964ApJ...140.1056O,2015MNRAS.446.2110T} and for magnetized models, see   \citet{2014ApJ...785...24T,2021ApJ...911..106K}.}.
Collision between two orthogonally running filaments leads to a monolithic collapse of the intersection when its mass is sufficient and the self-gravity is dominant \citep{2024ApJ...974..265K}. 

The star formation paradigm is now changing, paying attention to filaments.
The physical processes of filament-filament collision are now being studied both observationally and theoretically.
However its global effect is poorly known, e.g., Do collisions between filaments and those between clouds predict different star formation schemes?
Is star formation affected by the formation mechanisms of the filaments and clouds?
To answer such problems, it is necessary to construct a comprehensive model of star formation including the filament-filament collision. 
As part of this, we try to establish a semi-analytic model for filament-filament collisions.
In this paper, we focus on the collision cross section (CCS) of filament-filament collisions and explore the differences between them and cloud-cloud collisions.

The paper is structured as follows:
In section \ref{sec:2}, we derive a formulation for the CCS of two identical cylindrical filaments.
In calculating the CCS, we first handle thin filaments with negligible width. 
We then discuss the differences with finite-width filaments.
We show that the CCS of a filament depends on three vectors: the directions of the two filament axes and their relative velocity. 
In section \ref{sec:3}, we consider the distribution of the directions of the three vectors
 and discuss how the average CCS depends on the distribution function.
In section \ref{sec:4}, we discuss predictions from our model,
 the hub mass function expected from filament collisions,
 the relationship between the hub mass and the mass-dependent CCS,
 and the angle between the two filaments that form a massive hub.
We conclude in section \ref{sec:5}.
   
\section{Collision cross section (CCS)}\label{sec:2}

\subsection{Model}\label{ssec:2.1}
In this paper, we assume a cylindrical filament of length $2p$, radius $q$, and density $\rho_0$.   
Therefore, the line-mass of the filament is $\lambda_0=(\pi/2)q^2 \rho_0$, and the mass is $M_0=2\lambda_0 p$.
In contrast to the spherical cloud, the CCS of such a filament depends
 on the angles between the axes of the two filaments and their relative velocities.
We name the filaments as filament 1 and filament 2.
Let the directions of the two filaments be $\bm{n}_1$ and $\bm{n}_2$, respectively.
The direction of the relative velocity is defined by $\bm{n}_v\equiv\bm{v}/|\bm{v}|$, with the relative velocity being $\bm{v}=\dot{\bm{r}}_2-\dot{\bm{r}}_1$.
Here, $\bm{r}_1$ and $\bm{r}_2$ represent the position vectors of the centers of mass of filaments 1 and 2, respectively.
Filament $i$ (hereafter $i$ represents $i=1, 2$) extends from $\bm{r}_i - p\bm{n}_i$ to $\bm{r}_i + p\bm{n}_i$.

\subsection{Filaments with negligible width $q \ll p$}\label{ssec:2.2}
\begin{figure}[hbtp]
\begin{center}
\includegraphics[width=6cm]{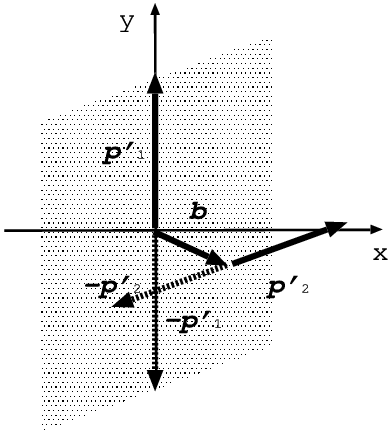}
\end{center}
\caption{Projection of filaments 1 and 2 onto a plane perpendicular to the relative velocity.
The impact parameter is denoted by $\bm{b}$.
The endpoints of filament 1 are $\bm{p}'_1$ and $-\bm{p}'_1$.
The endpoints of filament 2 are $\bm{b}+\bm{p}'_2$ and $\bm{b}-\bm{p}'_2$.
If the impact parameter $\bm{b}$ is selected inside the shaded parallelogram in the figure,
 the two filament vectors $\pm\bm{p}'_1$ and $\bm{b}\pm\bm{p}'_2$ intersect.
{Alt text: A schematic diagram which contains two directional vectors. 
We consider the condition in which these vectors intersect each other.} }
\label{fig.1}
\end{figure}

Similar to the procedure for calculating the CCS of a spherical cloud, 
 we ignore here the gravity working between the filaments.
Filaments are often found with an aspect ratio up to 10 [see figure 6(a) of \citet{2019A&A...621A..42A} and figure 10 of \citet{2023ASPC..534..153H}],
first, we consider a simple model with negligible width $q \ll p$.
Project vector $\bm{n}_i$ onto a plane perpendicular to the relative velocity.
The projection is obtained 
\begin{align}
\bm{n}'_1&=\bm{n}_1-(\bm{n}_1\cdot \bm{n}_v)\bm{n}_v,\label{eq:n1v}\\
\bm{n}'_2&=\bm{n}_2-(\bm{n}_2\cdot \bm{n}_v)\bm{n}_v.\label{eq:n2v}
\end{align} 
Hereafter, vectors with $'$ represent those projected onto a plane perpendicular to the relative velocity.
Because the filaments are represented by $\bm{p}_i=p\bm{n}_i$, 
 the collision condition is expressed using $\bm{p}'_i=p\bm{n}'_i$.
Figure \ref{fig.1} shows the arrangement of the filaments  projected onto the plane.
While maintaining the two vectors $\bm{p}'_i$ (directions and lengths), consider the allowable impact parameter $\bm{b}\equiv \bm{r}'_2-\bm{r}'_1$ for the two vectors to intersect each other.
If we select the impact parameter within the parallelogram shown in the figure,
 it is easy to see that the filaments will collide.
The CCS is equal to the area of the parallelogram and is expressed as follows: 
\begin{align}
S=\left|2\bm{p}'_1\times 2\bm{p}'_2\right|=4p^2\left|\bm{n}'_1\times \bm{n}'_2\right|. \label{eq:S}
\end{align}

The CCS depends on the orientations of the three vectors $\bm{n}_1$, $\bm{n}_2$, and $\bm{n}_v$ [equation (\ref{eq:S})]. 
It would be interesting to compare the average CCS of a filament and that of a spherical cloud.
If we name the angle between the filament $\bm{n}_i$ and the relative velocity $\bm{n}_v$ as $\theta_i$, 
 and the angle between the two projected vectors $\bm{n}'_i$ as $\phi_{12}$,
 then $|\bm{p}'_i|=|\sin\theta_i|\,p$ and $|\bm{n}'_1\times \bm{n}'_2|=|\bm{n}'_1||\bm{n}'_2||\sin\phi_{12}|$.
The average of $S$ is calculated as follows:
\begin{equation}
\overline{S}=4p^2\overline{\sin \theta}_1\, \overline{\sin \phi}_{12}\, \overline{\sin \theta}_2.
\end{equation}
If the orientations of the two filaments and their relative velocities all follow an isotropic distribution,
the average angle can be estimated as follows:
 $\overline{\sin\theta}_1=\overline{\sin\theta}_2=\int_0^{\pi/2}\sin^2\theta d\theta/\int_0^{\pi/2}\sin\theta d\theta=\pi/4$   
 and $\overline{\sin \phi}_{12}=\int_0^{\pi/2}\sin\phi d\phi/\int_0^{\pi/2}d\phi=2/\pi$,
 which leads to $\overline{S}=(\pi/2)p^2$.
Since the CCS between spherical clouds of radius $p$ is given by $S_\mathrm{sph}=4\pi p^2$,
 the average CCS between the filaments is 1/8 smaller than that of the spherical cloud.
  
\subsection{Filaments with a finite width $q\protect\lesssim p$}\label{ssec:2.3}

\begin{figure}[hbtp]
\begin{center}
\includegraphics[width=6cm]{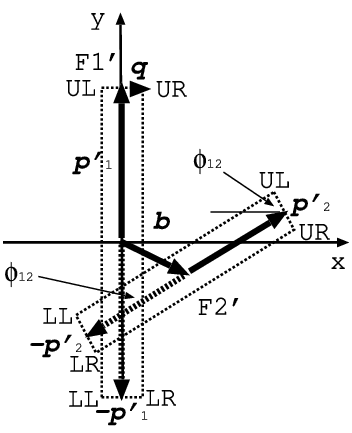}
\end{center}
\caption{Projection of finite-width filaments onto a plane perpendicular to the relative velocity (two dotted rectangles).
The centers of the filament ends are the same as in figure \ref{fig.1}. 
For filament 1, $\pm \bm{p}'_1$, for filament 2, $\bm{b}\pm\bm{p}'_2$.
The conditions for the two rectangles to overlap are considered.
See also appendix \ref{sec:A1}.
{Alt text: A schematic diagram which contains two rectangles.
We consider the condition in which these rectangles overlap each other.
Vertices of the rectangle are labeled UL, UR, LR, and LL clockwise from the top left.}
}
\label{fig.2}
\end{figure}
 
By projecting the cylindrical filament (3D) onto a plane perpendicular to the relative velocity,
 ignoring the shape of the filament ends, we obtain a rectangular (2D) with length $2|\bm{p}'_i|=2p|\bm{n}'_i|$
 and width $2q$ (figure \ref{fig.2}).
Taking the filament width into account, we derive the allowable region of the impact parameter $\bm{b}$
 and compare it with the result obtained previously when the width is negligible (the shaded parallelogram in figure \ref{fig.1}).
Appendix \ref{sec:A1} briefly explains the derivation of this condition,
 and figure \ref{fig.A1} shows the distribution of the impact parameter.     
The outer octagon indicates the region of the impact parameters required for two finite-width filaments to collide. 
The dark-shaded parts of figure \ref{fig.A1} are the same as those in figure \ref{fig.1}.

Figure \ref{fig.A1} in appendix \ref{sec:A1} shows that filaments of finite width have a larger CCS than those of negligible width.
Collision between two finite-width filaments occurs just when the point LL on filament 2 touches the line segment UR-LR on filament 1 (see figure \ref{fig.2}).
The terminus of $\bm{b}$ extends to the right by $q+\cos\phi_{12}q$ compared to the zero-width model.
Similarly, the terminus of $\bm{b}$ extends to the left by the same amount.
The CCS increases by $2(1+|\cos\phi_{12}|)q\cdot 2|\bm{p}'_{1}|$. 
Here, we change $\cos\phi_{12}$ to $|\cos\phi_{12}|$, 
 considering the fact that the CCS is preserved even when flipping $\bm{p}'_1$ to $-\bm{p}'_1$.

Considering the condition of whether point UL on filament 1 touches the line segment UR-LR on filament 2,
 we can see that $\bm{b}$ also extends upward by a width equal to $q+\cos\phi_{12}q$.
By symmetry,  $\bm{b}$ also extends downward by the same amount. 
From these two, the CCS increases by $2(1+|\cos\phi_{12}|)q\cdot 2|\bm{p}'_{2}|$.
As a result, the CCS increases by 
\begin{equation}
T=4(1+|\cos\phi_{12}|)q(|\bm{p}'_1|+|\bm{p}'_2|), \label{eq:T}
\end{equation}
and the CCS is given as $\Sigma=S+T$.
A more direct derivation is given in appendix \ref{sec:A1}.

The thin-shaded parallelogram in figure \ref{fig.A1} shares only the four long segments of the octagon,
 so $S+T$ represents the area of the parallelogram, which is different from that of the octagon. 
Because the difference between these two is small,
 the CCS for the two finite-width filaments to collide can be expressed as follows:
\begin{equation}
\Sigma=4\left|\bm{n}'_1\times \bm{n}'_2\right|p^2+4(1+|\cos\phi_{12}|)(|\bm{n}'_1|+|\bm{n}'_2|)pq+\mathcal{O}(q^2),
\label{eq:CCS}
\end{equation}
 [see also equation (\ref{eq:A1CCS})].
 
We compare $S$ and $T$ by averaging the equation for the isotropic distribution of filament direction and relative velocity.
Since $(1+\overline{|\cos\phi_{12}|})(\overline{|\sin\theta_1|}+\overline{|\sin\theta_2|})=(1+2/\pi)(\pi/2)$, we obtain $\overline{T}=4(1+\pi/2)pq\simeq 10.28pq$.
Hence, $\overline{T}/\overline{S}=8(1+\pi/2)/\pi\times (q/p)\simeq 0.6546 [(q/p)/0.1]$.
If the axial ratio of the observed filament is $q/p\lesssim 0.1$ , then $\overline{T}/\overline{S}\lesssim 0.5$.
Therefore, 
\begin{equation}
\overline{\Sigma}=[\pi/2+4(1+\pi/2)(q/p)]p^2\simeq \{1.57+1.03[(q/p)/0.1]\}p^2,
\end{equation}
 which is 0.2 times smaller than the CCS of the spherical cloud.
 
\section{Confined distributions} \label{sec:3}

In the preceding section (subsections \ref{ssec:2.2} and \ref{ssec:2.3}),
 we calculated the CCS averaged over an isotropic distribution of $\bm{n}_i$ and $\bm{n}_v$. 
However, filaments are more likely to form from disk-like structures (e.g., \cite{2013ApJ...774L..31I,2021ApJ...916...83A}).
For such a formation mechanism, anisotropic distributions would be expected.
Here, we investigate how these anisotropic distributions affect the CCS.   

For simplicity, we use the variable $\alpha$ to represent the three kinds of indices $\alpha=1, 2$, and $v$.
Spherical coordinates of the three vectors  $\bm{n}_\alpha$ ($\alpha=1, 2$, and $v$)
 are determined by the six variables, $\theta_\alpha$ and $\phi_\alpha$.
By converting the zenith angle $\theta$ to $\mu\equiv\cos\theta$,
 the distribution is specified by the distribution function $f_\alpha(\mu_\alpha,\phi_\alpha)$.
The average CCS is calculated as
\begin{align}
\overline{\Sigma}=&\int_{\mu_1=-1}^1\int_{\mu_2=-1}^1\int_{\mu_v=-1}^1
\int_{\phi_1=0}^{2\pi}\int_{\phi_2=0}^{2\pi}\int_{\phi_v=0}^{2\pi}\nonumber \\
&\Sigma(\mu_1,\mu_2,\mu_v,\phi_1,\phi_2,\phi_v)\nonumber \\
&f_1(\mu_1,\phi_1)f_2(\mu_2,\phi_2)f_v(\mu_v,\phi_v) d\mu_1 d\mu_2 d\mu_v d\phi_1 d\phi_2 d\phi_v.
\end{align}

For an isotropic distribution of $\bm{n}_\alpha$, the mean is calculated using
\begin{align}
&f_\alpha(\mu_\alpha, \phi_\alpha)d\mu_\alpha d\phi_\alpha=\frac{d\mu_\alpha}{2}\frac{d\phi_\alpha}{2\pi}.
\label{eq:distribution_function_isotropic}
\end{align}
For the grid sizes $\Delta\mu_\alpha=2/40$ and $\Delta\phi_\alpha=2\pi/40$,
 the trapezoidal integration yields $\overline{S} \simeq 1.57p^2$ and $\overline{T}\simeq 10.27pq$, 
 which agrees with the analytically obtained values $(\pi/2)p^2=1.570796...p^2$ and $4(1+\pi/2)pq=10.283185...pq$, respectively.
Table \ref{tbl.1} summarizes the distribution function choices.
Model I corresponds to this isotropic model, where the distribution functions of both $\bm{n}_i$ and $\bm{n}_v$
 are taken as shown in equation (\ref{eq:distribution_function_isotropic}).
  
\subsection{Models with uniform distribution in $\phi$'s} \label{ssec:3.1}

\begin{figure}[hbtp]
\begin{center}
\includegraphics[width=8cm]{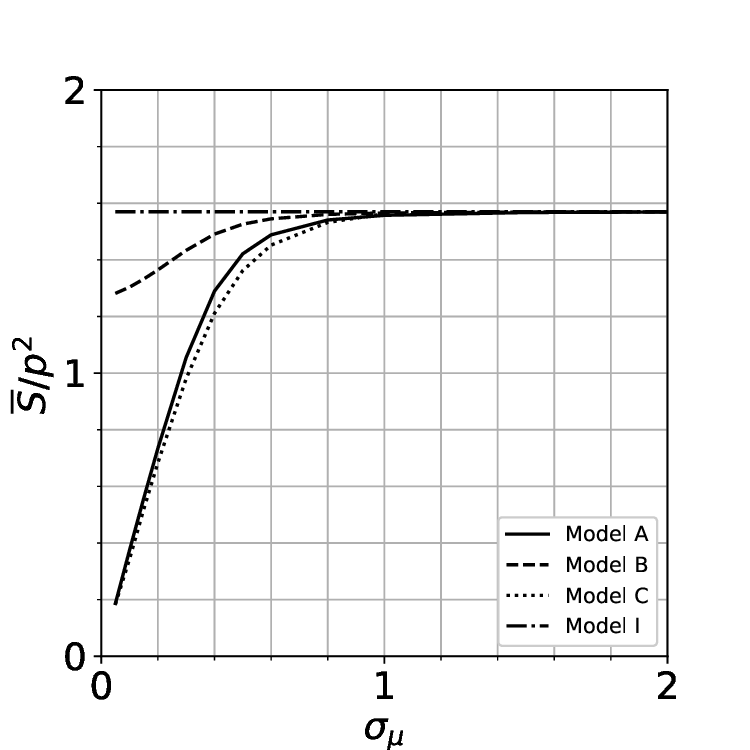}
\end{center}
\caption{\label{fig.3}
The mean $\overline{S}$ for various distributions of $\bm{n}_\alpha$.
We calculate four models for different sets of distribution functions, as shown in table \ref{tbl.1}.
The more concentrated the distribution functions we choose,
 the more significantly the mean $\overline{S}$ decreases in the range $\sigma_\mu\lesssim 1$. 
 {Alt text: A four-line graph.  
 Models A, B, C, and I are shown.
 Horizontal and vertical axes represent sigma sub mu
 and S bar over p squared, respectively.}
}
\end{figure}

\begin{table}[htp]
\caption{Model distribution function.}
{\tabcolsep = 2pt
\begin{tabular}{llll}
Model & $(\mu_i, \phi_i)$ &  $(\mu_v, \phi_v)$ & Distribution\\
\hline\hline
A\footnotemark[$*$]\ldots\ldots & eq.(\ref{eq:f_normal_mu+uniform_phi})  &eq.(\ref{eq:f_normal_mu+uniform_phi}) & $\bm{n}_\alpha$: TND\footnotemark[$\#$] in $\mu_\alpha$\\
B\footnotemark[$\dag$]\ldots\ldots & eq.(\ref{eq:f_normal_mu+uniform_phi}) &eq.(\ref{eq:distribution_function_isotropic}) & $\bm{n}_i$: TND in $\mu_i$; $\bm{n}_v$: isotropic \\
C\footnotemark[$\ddag$]\ldots\ldots & eq.(\ref{eq:f_normal_mu+normal_phi}) &eq.(\ref{eq:distribution_function_isotropic}) & $\bm{n}_i$: TND in $\mu_i\,\&\,\phi_i$; $\bm{n}_v$: isotropic\\
I\footnotemark[$\S$] \ldots\ldots  & eq.(\ref{eq:distribution_function_isotropic}) & eq.(\ref{eq:distribution_function_isotropic}) & $\bm{n}_\alpha$: isotropic distribution \\
\hline
\end{tabular}}
\label{tbl.1}
\begin{tabnote}
\footnotemark[$*$] Solid line in figure \ref{fig.3}.\\
\footnotemark[$\dag$] Dashed line.\\
\footnotemark[$\ddag$] Dotted line.\\
\footnotemark[$\S$] Dash-dot line.\\
\footnotemark[$\#$] Truncated normal distribution.
\end{tabnote}
\end{table}%

We consider the following model in which the distribution function is concentrated near the $\mu=0$ plane but uniform within $\phi$ as
\begin{align}
f_\alpha(\mu_\alpha, \phi_\alpha)d\mu_\alpha d\phi_\alpha=&
\frac{\exp\left(-{\mu^2_\alpha}/{2\sigma^2_{\mu_\alpha}}\right)d\mu_\alpha}{\left(2\pi\sigma^2_{\mu_\alpha}\right)^{1/2}\mathrm{Erf}\left[\left(2\sigma^2_{\mu_\alpha}\right)^{-1/2}\right]}
\frac{d\phi_\alpha}{2\pi}, \nonumber\\
&\ \ \ (-1\le\mu_\alpha\le 1)
\label{eq:f_normal_mu+uniform_phi}
\end{align}
where the distribution of $\mu_\alpha$ is a truncated normal distribution (TND), and $\sigma_{\mu_\alpha}$ denotes the standard deviation of $\mu_\alpha$ around the mean $\overline{\mu}_\alpha=0$.
Here, $\mathrm{Erf}(x)$ represents the Gauss error function defined as
$\mathrm{Erf}(x)=2\pi^{-1/2}\int_0^x\exp(-t^2)dt$.

Figure \ref{fig.3} shows the mean $\overline{S}$ for various $\bm{n}_\alpha$ distributions.  
The solid line (Model A) represents the CCS corresponding to the distribution of equation (\ref{eq:f_normal_mu+uniform_phi}).
When $\sigma_\mu \gtrsim 1$, the mean CCS coincides with that of an isotropic distribution ($\overline{S}/p^2\simeq 1.57$; dash-dotted line).
By decreasing $\sigma_\mu\lesssim 1$, the CCS decreases to 1/2 of that of the fully isotropic model when the standard deviation becomes as small as $\sigma_{\mu_\alpha}\simeq 0.2$.
That is, if $\bm{n}_i$ prefers a disk orientation with $1\sigma=\pm 12\mathrm{deg}$, the CCS $\overline{S}$ will decrease to this value.

The dashed line (Model B) represents a model with $\sigma_{\mu_i}=\sigma_\mu$ and $\sigma_{\mu_v}=10$.
In this model, the filament directions $\bm{n}_i$ follow the distribution given by equation (\ref{eq:f_normal_mu+uniform_phi}),
 but the relative velocities $\bm{n}_v$ are isotropically distributed\footnote{Taking into account of $|\mu_v|\le 1$, a large standard deviation means that the distribution is uniform in $\mu_v$.}.
In this model, the reduction from the CCS for the isotropic distribution is smaller than that for the distribution in Model A.   
In Models A and B, $\bm{n}_i$ are concentrated toward $\mu_i=0$. 
When also $\bm{n}_v$ has a similar distribution (Model A), reduction in the CCS $S$ becomes more prominent 
 for small $\sigma_\alpha\lesssim 1$, in comparison with Model B with the isotropic velocity distribution.
This indicates that the more restricted the relative velocity distribution, the more pronounced the reduction of the CCS $S$ becomes.  

\subsection{Model with alignment also in $\phi$ direction} \label{ssec:3.2}

In the preceding subsection \ref{ssec:3.1}, we considered a model in which the filament axis is aligned with a disk.
Because the interstellar magnetic field is thought to determine the direction of the filament, 
 here we consider a model in which the filament axis points in a specific direction in $\phi$.
We consider the distribution function of the filament axis as
\begin{align}
&f_i(\mu_i, \phi_i)d\mu_i d\phi_i=
\frac{\exp\left(-{\mu^2_i}/{2\sigma^2_{\mu_i}}\right)d\mu_i}{\left(2\pi\sigma^2_{\mu_i}\right)^{1/2}\mathrm{Erf}\left[\left(2\sigma^2_{\mu_i}\right)^{-1/2}\right]}\nonumber\\
&\times\frac{\exp\left\{-{\mathrm{min}\left[\left(\phi_i-\pi/2\right)^2,\left(\phi_i-3\pi/2\right)^2\right]}/{2\sigma^2_{\phi_i}}\right\}d\phi}
{2(2\pi\sigma^2_{\phi_i} )^{1/2}\mathrm{Erf}\left[(\pi/2)(2\sigma^2_{\phi_i})^{-1/2}\right]}.\nonumber\\ 
&\ \ \ \ \ \ \ (-1\le\mu_i\le 1,\ 0\le \phi_i\le 2\pi)
\label{eq:f_normal_mu+normal_phi}
\end{align}
Here, the distribution function is composed of two truncated normal distributions.
The distribution in the $\mu_i$ direction is the same as equation (\ref{eq:f_normal_mu+uniform_phi}).
For the $\phi_i$ direction, we consider a distribution with a mean and standard deviation of
 $\overline{\phi}_i=\pi/2$ or $3\pi/2$ and $\sigma_i$. 
That is,  $\phi_i=\pi/2\pm \sigma_i$ or $\phi_i=3\pi/2\pm \sigma_i$.
To make the variance of $\phi_i$ equivalent to that obtained from the standard deviation $\sigma_\mu$, 
  we take $\sigma_\phi$ to be
\begin{equation}
\sigma_{\phi_i}=\mathrm{arcsin}\, \sigma_{\mu_i}. \ \ \ \ (0 \le \sigma_{\mu_i} \le 1)
\end{equation}
In this model, both filaments are aligned with the $y$-axis.
The degree of alignment is controlled by the parameter $\sigma_\mu$.
The dotted line in figure \ref{fig.3} represents this model consisting of two types of distribution functions:
 equation (\ref{eq:f_normal_mu+normal_phi}) for $i=1, 2$
 and equation (\ref{eq:distribution_function_isotropic}) for $\alpha=v$, which is called Model C.   
 
Using Model B as the base [equations (\ref{eq:f_normal_mu+uniform_phi}) and (\ref{eq:distribution_function_isotropic})],
 Model C assumes a more restricted distribution for $\bm{n}_i$ [equations (\ref{eq:f_normal_mu+normal_phi}) and (\ref{eq:distribution_function_isotropic})], resulting in a significant decrease in the mean CCS for small $\sigma_{\mu_i}$.
Model A assumes a more restricted distribution for $\bm{n}_v$ [equations (\ref{eq:f_normal_mu+uniform_phi}) and (\ref{eq:f_normal_mu+uniform_phi})], the mean CCS decreases significantly for small $\sigma_{\mu_v}$.
Model I assumes less restrictive distributions for both $\bm{n}_i$ and $\bm{n}_v$ [equations (\ref{eq:distribution_function_isotropic}) and (\ref{eq:distribution_function_isotropic})], and therefore by definition no decrease can be observed.
In summary, the decrease in CCS becomes more pronounced when we assume a better alignment of the three vectors $\bm{n}_\alpha$.

\subsubsection{$S$ and $T$} 

\begin{figure}[htbp]
\begin{center}
\includegraphics[width=8cm]{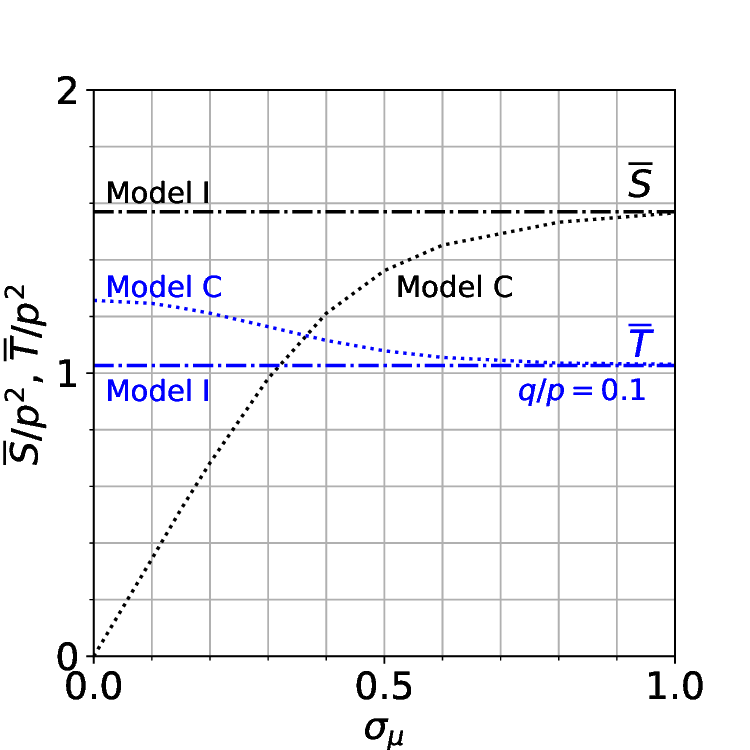}
\end{center}
\caption{Comparison of $\overline{S}$ and $\overline{T}$.
For the distribution function of Model C (table \ref{tbl.1}), 
the leading term of the CCS, $\overline{S}$ (black dotted line), proportional to $p^2$,
 and the second term of the CCS, $\overline{T}$ (blue dotted line), proportional to $pq$, are plotted
 against the standard deviations of the truncated normal distribution $\sigma_\mu$ and $\sigma_\phi$.
The filament axis is assumed to follow the distribution specified by
$\sigma_{\mu}=\sigma_{\mu_1}=\sigma_{\mu_2}$ and
$\sigma_{\phi_1}=\sigma_{\phi_2}=\arcsin\sigma_{\mu}$, which tends to align toward the $y$-axis.
The relative velocity field is assumed to be isotropic.
When all distributions are isotropic (Model I), the means $\overline{S}$ (black dash-dotted line) and $\overline{T}$ (blue dash-dotted line) are plotted as horizontal lines.
For comparison, we assume that the axial ratio is $q/p=0.1$. 
{Alt text: Four-line figure.
Horizontal axis represents sigma sub mu and vertical axis represents S bar over p squared and T bar over p squared.}
}
\label{fig.4}
\end{figure}

In the preceding section, we examined the mean $\overline{S}$ (CCS proportional to $p^2$)
 for various distributions in the three axial directions.
Next, we compare this with $\overline{T}$, which is the CCS proportional to $pq$ [equation (\ref{eq:T})].
The distribution function is Model C and the axis ratio is $q/p=0.1$.
Figure \ref{fig.4} shows the changes in $\overline{S}$ and $\overline{T}$ with respect to $\sigma_\mu$.
When we assume $\sigma_\mu\ll 1$, the truncated normal distribution of equation (\ref{eq:f_normal_mu+normal_phi})
 is reduced to the Dirac's delta function as 
\begin{equation}
f_i(\mu_i,\phi_i) d\mu_i d\phi_i=\delta(\mu_i)\delta(\min(|\phi_i-\pi/2|,|\phi_i-3\pi/2|) d\mu_i d\phi_i.
\end{equation}
Then, the average CCSs of $S$ [equation (\ref{eq:S})] and $T$ [equation (\ref{eq:T})] are given as
$\overline{S}(\sigma_\mu\rightarrow 0)=0$ and 
$\overline{T}(\sigma_\mu\rightarrow 0)=4\pi pq\simeq 12.57pq$, respectively.

For the average CCS of an isotropic distribution, $\overline{S}$ is larger than $\overline{T}$ (figure \ref{fig.4}).
For the Model C distribution functions, $\overline{S}$ is larger than $\overline{T}$ when $\sigma_\mu\gtrsim 0.4$.
However, as $\sigma_\mu$ decreases, $\overline{S}$ decreases and $\overline{T}$ increases.
From the comparison between the first two terms of equation (\ref{eq:CCS}),  this behavior can be understood:
When the two vectors $\bm{n}_i$ become parallel, the angle between the projected vectors $\bm{n}'_i$ decreases,
 becoming $\sin\phi_{12}\rightarrow 0$, while $\cos\phi_{12}\rightarrow 1$.
In other words, as $\sigma_{\mu_i}$ decreases and the two filaments align along the $y$-axis,
 $\overline{S}$ decreases, but $\overline{T}$ does not.

\section{Predictions from the model}\label{sec:4}

\subsection{Mass function of hubs}\label{ssec:4.1}
The CCS provides a clue for deriving the hub mass function.
The mass function depends on several factors.
Using the probability of a state specified by the three vectors $\bm{n}_\alpha$, $f_\alpha(\bm{n}_\alpha)$,
 the probability of such a collision is proportional to $\Sigma(\bm{n}_\alpha)f_\alpha(\bm{n}_\alpha)$.
If the hub mass expected from the collision configuration is given as $M_\mathrm{hub}(\bm{n}_\alpha)$,  
 the mass function can be obtained as the relationship between $M_\mathrm{hub}(\bm{n}_\alpha)$
 and $A\Sigma(\bm{n}_\alpha)f_\alpha(\bm{n}_\alpha)$.
  
\begin{figure}[htbp]
{\centering
\includegraphics[width=8cm,bb=0 0 523 508]{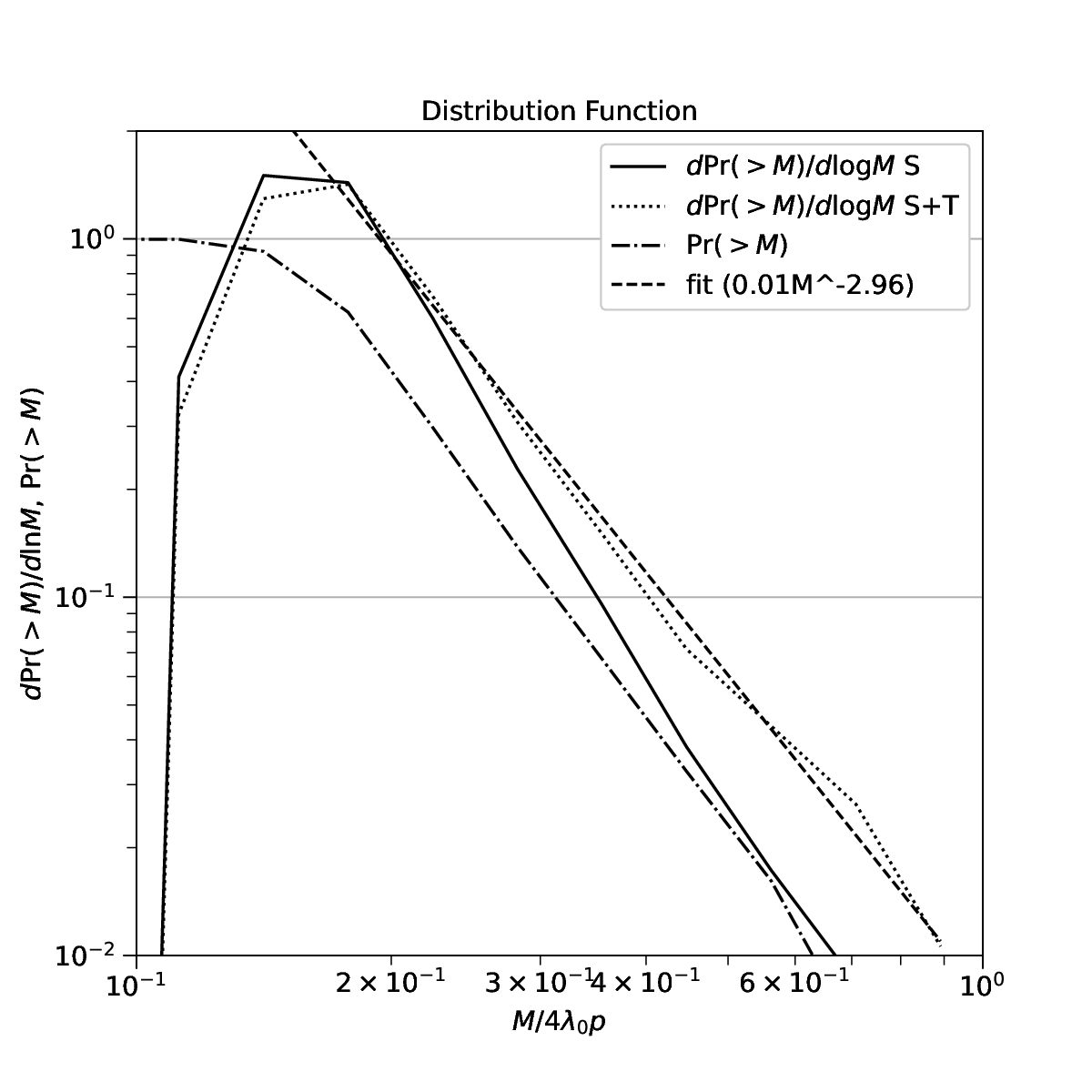}
}
\caption{Mass function (MF) of the hub expected from the filament-filament collisions.
MF is plotted against the hub's mass normalized by the total mass of the two filaments $2M_0=4\lambda_0 p$.
$d\mathrm{Pr}(M_\mathrm{Hub}>M)/d\log M$ for the CCS of $\Sigma=S$ (solid line)
and that for $\Sigma=S+T$ (dotted line) are plotted.
When calculating $T$, we assume the axis ratio $q/p=0.1$. 
The power-law fitting for $\Sigma=S+T$ model is also shown by a dashed line.
{Alt text: Four-line figure.
Horizontal axis represents the hub mass divided over 4  lambda sub 0 p.
Vertical axis represents the mass function.}
}
\label{fig.5}
\end{figure}

When two filaments collide, the overlapping portion of the two filaments forms a hub,
 while the remaining portions continue to move as filaments.
The exact amount of the hub's mass must be obtained from numerical simulations \citep{2024ApJ...974..265K}.
However, here we employ a simple formula to derive the overlapping mass when two identical cylindrical filaments collide in a  configuration 
 specified with $\bm{n}_\alpha$.
As shown in appendix \ref{sec:A2},
 the mass of filament 1 swept by filament 2 is given by $\Delta M_1=2q\lambda_0/\cos\beta_1$ [equation (\ref{eq:DM1})].
Here, $q$, $\lambda_0$, and $\beta_1$ represent the filament radius, its line-mass,
 and the angle between the vectors $\bm{n}_v\times \bm{n}_2$ and $\bm{n}_1$, respectively. 
Since $\Delta M_1$ is limited by the mass of one filament, $M_0=2p\lambda_0$, 
 for simplicity, we assume that the mass supplied to the hub from filament 1 is limited as
  $\Delta M_1=\mathrm{Min}(2q\lambda_0/\cos\beta_1, 2p\lambda_0)$.  
A similar mass is supplied from filament 2, giving $\Delta M_2=\mathrm{Min}(2q\lambda_0/\cos\beta_2, 2p\lambda_0)$ [see equation (\ref{eq:DM2})].
Here, $\beta_2$ represents the angle between vector $\bm{n}_1\times \bm{n}_v$ and vector $\bm{n}_2$.
Therefore, the overlapping mass is expressed as follows:
\begin{align}
M_\mathrm{hub}&\equiv \Delta M_1+\Delta M_2 \nonumber\\
&=4\lambda_0 p\frac{1}{2}\left\{\mathrm{Min}\left[\frac{q/p}{\cos\beta_1(\bm{n}_\alpha)}, 1\right]
                             +\mathrm{Min}\left[\frac{q/p}{\cos\beta_2(\bm{n}_\alpha)}, 1\right] \right\}.\label{eq:Mhub}
\end{align}
Considering the orthogonal collision $\bm{n}_1\perp\bm{n}_2\perp\bm{n}_v$,
 equation (\ref{eq:Mhub}) gives $M_\mathrm{hub}=4q\lambda_0$, which closely matches the numerically obtained mass  \citep{2024ApJ...974..265K}. 
Since equation (\ref{eq:Mhub}) gives the minimum and maximum of $M_\mathrm{hub}$ as
 $4\lambda_0 q \le M_\mathrm{hub} \le 4\lambda_0 p$, the expected mass range is within $1:p/q$ ratio.
Thus, the axis ratio of the filaments approximately describes the mass range of the hubs produced by the collision.  

Here, we simply assume that the distribution of $\bm{n}_\alpha$ is isotropic and
 that the probability of selecting $\bm{n}_\alpha$ 
 [i.e., the six angles are within the respective ranges of 
 $\theta_1-(\theta_1+\Delta\theta_1)$, $\phi_1-(\phi_1+\Delta\phi_1)$,
 $\theta_2-(\theta_2+\Delta\theta_2)$, $\phi_2-(\phi_2+\Delta\phi_2)$,
 $\theta_v-(\theta_v+\Delta\theta_v)$, and $\phi_v-(\phi_v+\Delta\phi_v)$]
  is proportional to
\begin{align}
\Delta f&=\sin\theta_1 \Delta \theta_1 \Delta \phi_1  \sin\theta_2 \Delta \theta_2 \Delta \phi_2 \sin\theta_v \Delta \theta_v \Delta \phi_v\nonumber\\
&=\Delta\mu_1\Delta \phi_1\Delta\mu_2\Delta \phi_2\Delta\mu_v\Delta \phi_v.    
\end{align}
If we select a parameter set such that $M_\mathrm{hub}>M$,
 the resulting probability of a collision in which the hub mass $M_\mathrm{hub}$ exceeds $M$ is given by
\begin{align}
\mathrm{Pr}(M_\mathrm{hub}>M)=\frac{\displaystyle\sum_{M_\mathrm{hub}>M}(S+T) \Delta f}{\displaystyle\sum_\mathrm{all}(S+T) \Delta f}=\frac{\displaystyle\sum_{M_\mathrm{hub}>M}(S+T)}{\displaystyle\sum_\mathrm{all}(S+T)}.
\end{align}
In the last part, we assume that all the independent variables $\mu_\alpha$ and $\phi_\alpha$ are equally divided in numerical integration, so that the term $\Delta f$ has a common constant value for each grid point.
The hub mass function is obtained as the relation between $M$ and $\mathrm{Pr}(M_\mathrm{hub}>M)$.

Figure \ref{fig.5} plots the probability distribution functions per unit $\Delta \log M$, $d\mathrm{Pr}(M_\mathrm{Hub}>M)/d\log M=M\,d\mathrm{Pr}(M_\mathrm{Hub}>M)/dM$, 
 for the two models, $\Sigma=S$ (solid line) and $\Sigma=S+T$ (dotted line).
Both mass functions (MFs) are well fitted by the power-law functions:
 $d\mathrm{Pr}(M_\mathrm{Hub}>M)/d\log M\propto M^{-3.78}$ ($\Sigma=S$) and $\propto M^{-2.96}$ ($\Sigma=S+T$).
Comparing these two, the $\Sigma=S$ model has a steeper slope,
 indicating that the CCS $T$ contributes more to the formation of the massive hubs through collisions.
This is natural when considering two parallel filaments colliding with perpendicular relative velocities, 
 $\bm{n}_1\parallel \bm{n}_2 \perp \bm{n}_v$,
 which induces a relatively large mass hub of $M_\mathrm{hub}\lesssim 4\lambda_0p=2M_0$.  
 
When $q/p=0.1$, the power-law exponent $\gamma_M$ of $\Sigma=S+T$ is $\gamma_M=-2.96$.
This becomes $\gamma_M=-2.91$ when $q/p=0.05$.
Thus, the power exponent seems almost independent of the parameter $q/p$.
 
The cumulative probability distribution $\mathrm{Pr}(M_\mathrm{Hub}>M)$ (dash-dotted line of figure \ref{fig.5}) indicates that collisions forming massive hubs exceeding half the total mass of the two filaments occur
in $\sim 3\%$ of all collisions.    
The above probability is for the isotropic distribution of $\bm{n}_\alpha$.
When considering a situation where the filaments are properly aligned
 [i.e., $\sigma_{\mu_i}$ and $\sigma_{\phi_i}$ are chosen small in equation (\ref{eq:f_normal_mu+normal_phi})],
 collisions with relatively large $M_\mathrm{hub}$ are selected,
 and the above probability seems to increase with the degree of alignment.  

\subsection{Mass-dependent CCS}
\begin{figure}[htbp]
\begin{center} 
\includegraphics[width=8cm]{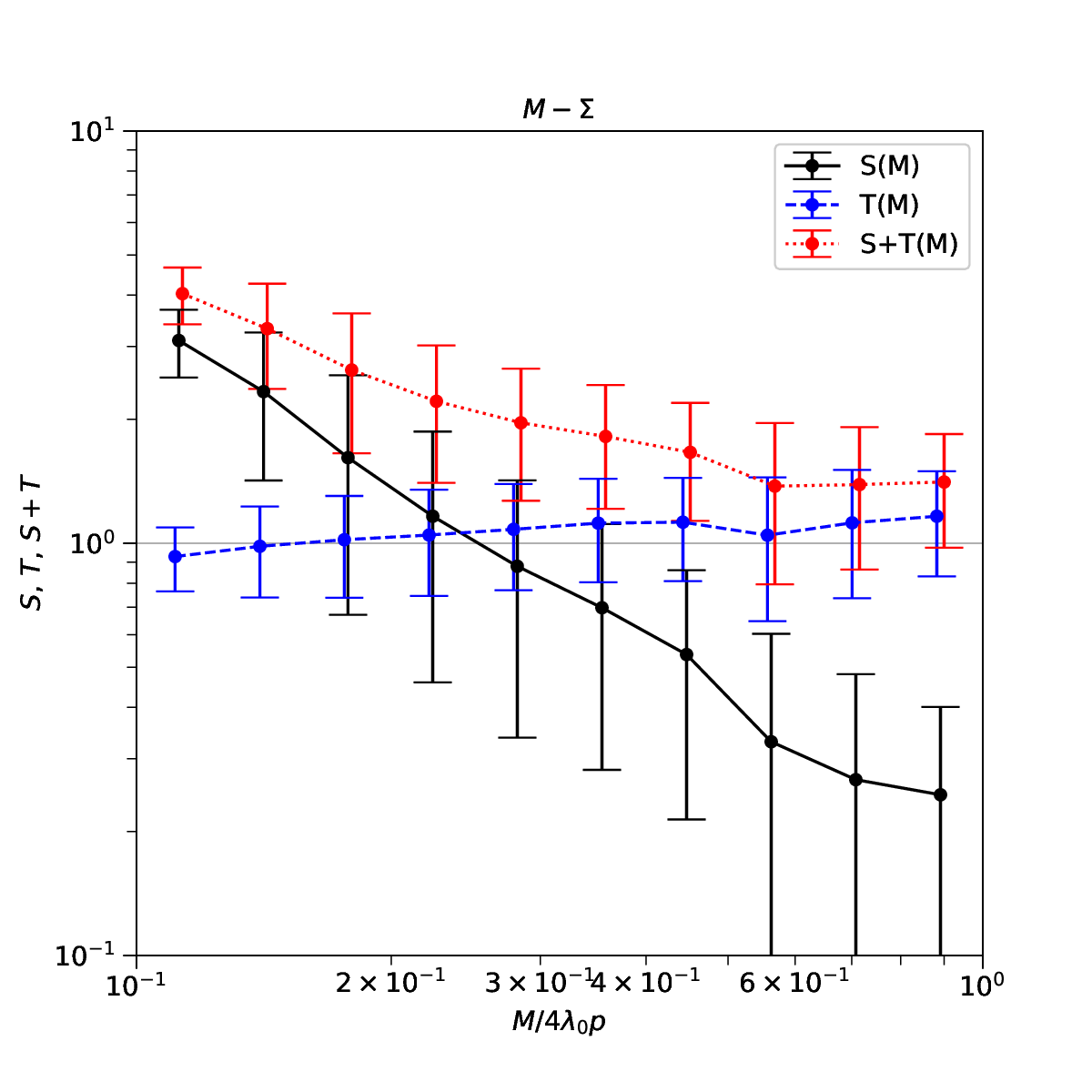}
\end{center}
\caption{Means and standard deviations of the mass-dependent CCSs $S$ (solid line), $T$ (dashed line), and $S+T$ (dotted line).
Means and standard deviations were calculated for each mass bin.
The vertical bars represent standard deviations.
{Alt text: Three-line figure.
Horizontal axis represents M over 4 lambda sub 0 p.
Vertical axis represents the collisional cross sections.}
}
\label{fig.6}
\end{figure}
 
Figure \ref{fig.6} plots the CCS for each mass bin $M-(M+\Delta M)$.
The mass-dependent CCS is calculated as follows:
For all configurations of $\bm{n}_\alpha$ ($\alpha=1, 2, v$),
 we calculate $S(\bm{n}_\alpha)$, $T(\bm{n}_\alpha)$, 
 and $M_\mathrm{hub}(\bm{n}_\alpha)$, using equations (\ref{eq:S}), (\ref{eq:T}), and (\ref{eq:Mhub}), respectively.
Figure \ref{fig.6} is obtained by calculating the average collision cross section in each mass bin. 
The mass bin is taken as $\Delta \log_{10} M=0.1$ or $\Delta M=10^{0.1}M$.

Comparing $\Sigma=S$ (solid line) and $\Sigma=T$ (dashed line),
 we see that as $M_\mathrm{hub}$ increases, $\Sigma=S(M_\mathrm{hub})$ decreases,
 while $\Sigma=T$ remains roughly constant.
As a result, the decrease in the total CCS, $\Sigma=S+T$ (dotted line), is smaller than $\Sigma=S$ (solid line). 
This qualitatively explains the slope of the MF.
Since the MF is proportional to the CCS, the MF calculated for $\Sigma=S$ shows a steeper slope than that for $\Sigma=S+T$ (figure \ref{fig.5}).    
\begin{figure}[htbp]
\begin{center}
(a)\hspace*{3.7cm}(b)\\
\includegraphics[width=4cm]{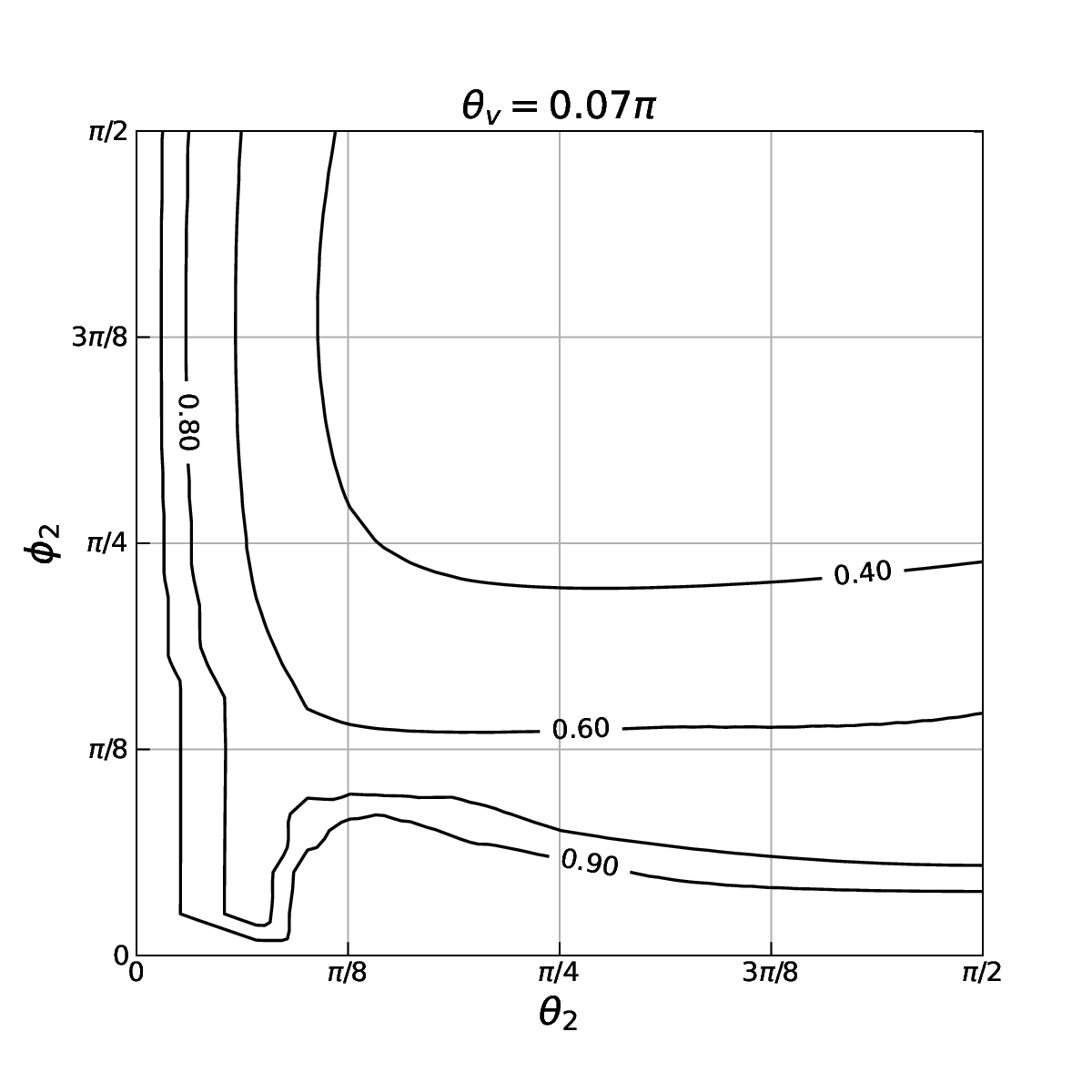}
\includegraphics[width=4cm]{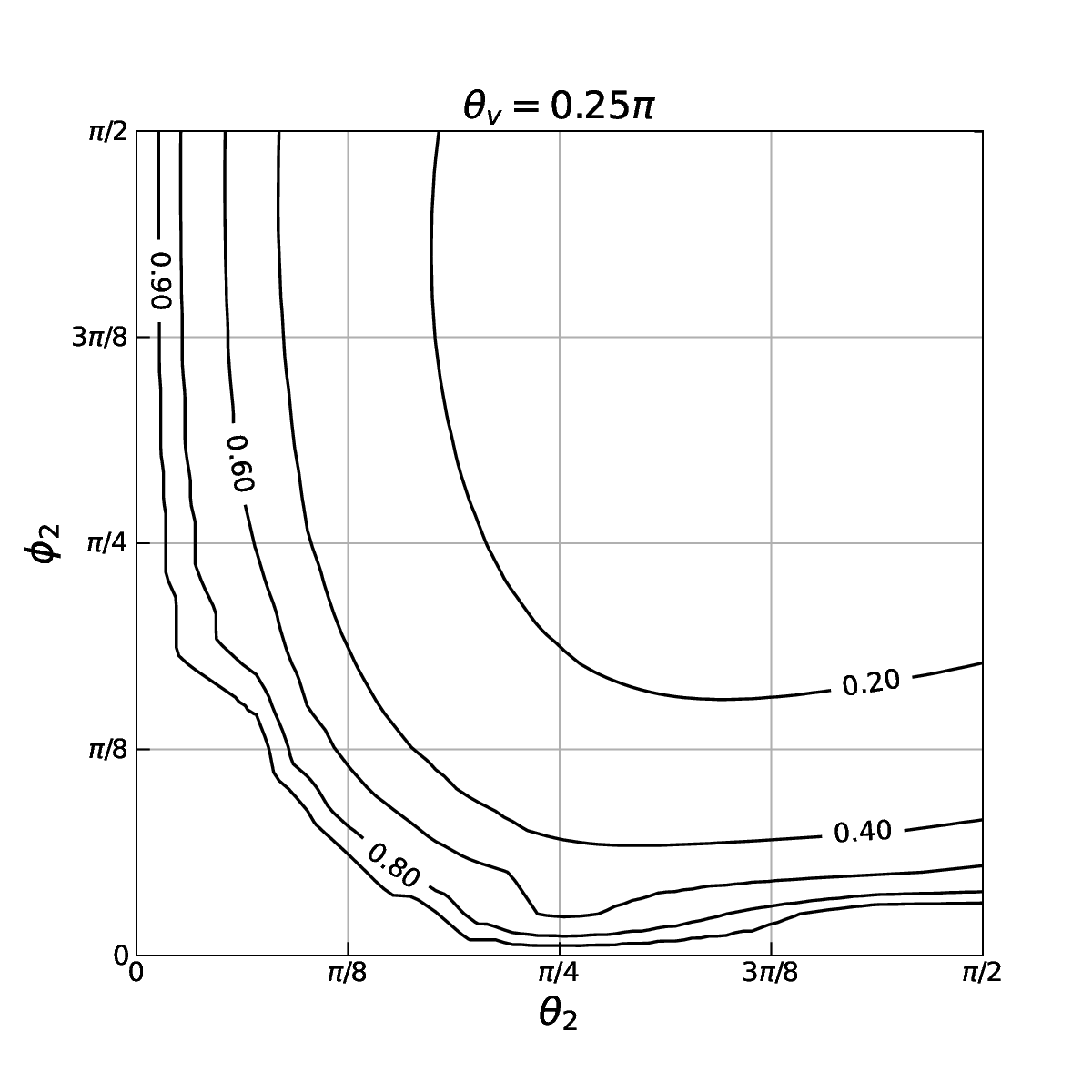}\\
(c)\\
\includegraphics[width=4cm]{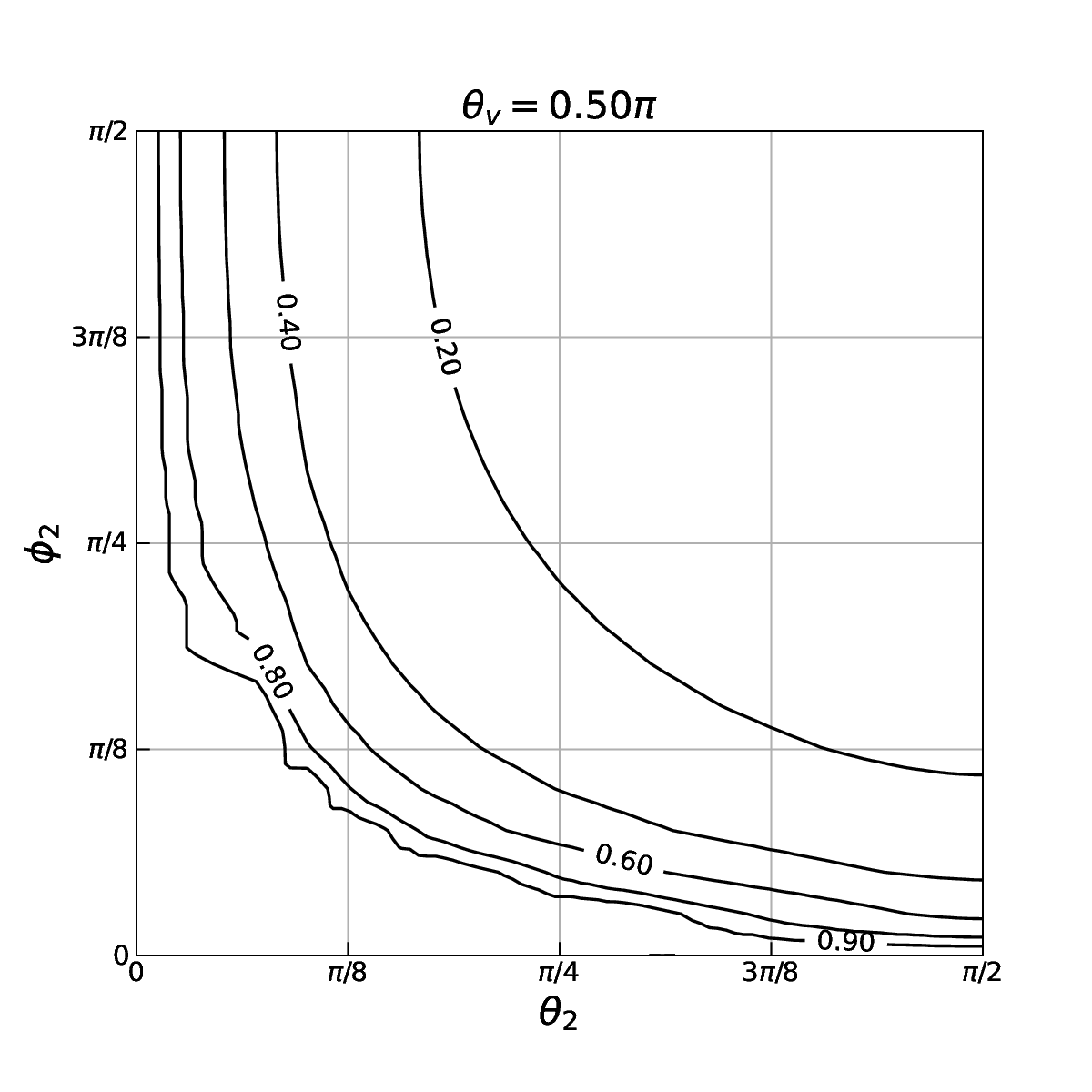}
\end{center}
\caption{
Hub mass expected for the direction of filament 2, $\bm{n}_2(\theta_2,\phi_2)$.
The orientations of other two vectors are taken $\bm{n}_1={}^t(0,0,1)$,
 and $\bm{n}_v={}^t(\sin\theta_v,0,\cos\theta_v)$, respectively. 
Panels (a), (b), and (c) correspond to $\theta_v\ll 1$
, $\theta_v=\pi/4$, and $\theta_v=\pi/2$, respectively. 
We plot $F\equiv M_\mathrm{hub}/(4\lambda_0p)$ using equation (\ref{eq:Mhub}). 
The axis ratio is $q/p=0.1$.
{Alt text: Three contour plots. 
Horizontal axis is theta sub 2 and vertical axis is phi sub 2.}
}
\label{fig.7}
\end{figure}

\subsection{Collisions resulting in massive overlap masses}

Equation (\ref{eq:Mhub}) shows that the mass of the hub produced by a collision between filaments
 depends on the orientation of the three vectors $\bm{n}_1$, $\bm{n}_2$, and $\bm{n}_v$.
What are the conditions for forming massive hubs that help massive star formation \citep{2003ApJ...585..850M}? 

Figure \ref{fig.7} shows how $M_\mathrm{hub}$ depends on the orientation of filament 2, $\bm{n}_2$, 
 if other two vectors are fixed as $\bm{n}_1={}^t(0,0,1)$ and $\bm{n}_v={}^t(\sin\theta_v,0,\cos\theta_v)$.
In figure \ref{fig.7}, we plot  $M_\mathrm{hub}$ [equation (\ref{eq:Mhub})] normalized by the total filament mass $4p\lambda_0$ as
\begin{equation}
F\equiv \frac{M_\mathrm{hub}}{4p\lambda_0}=\frac{1}{2}\left\{\mathrm{Min}\left[\frac{q/p}{\cos\beta_1(\bm{n}_\alpha)}, 1\right]
                             +\mathrm{Min}\left[\frac{q/p}{\cos\beta_2(\bm{n}_\alpha)}, 1\right] \right\}.\label{eq:F}
\end{equation}
By varying the three angles $\theta_v$, $\theta_2$, and $\phi_2$,
 we can see the angular conditions that result in large overlapping masses. 

Figure \ref{fig.7} shows that for all $\theta_v$, $M_\mathrm{hub}$ reaches a maximum  ($F\rightarrow 1$),
 as it approaches the horizontal and vertical axes, i.e.,  $\phi_2\rightarrow 0$ or $\theta_2\rightarrow 0$.
Furthermore, comparing panel (a) with panels (b) and (c), 
 we see that collisions with $\theta_v\ll 1$ result in larger $F$ than the cases for $\theta_v=\pi/4$ and $\theta_v=\pi/2$. 
Figure \ref{fig.7} shows that the overlapping mass increases for three special configurations of the three vectors:
\begin{enumerate}
\item 
Mass $F$ increases with decreasing $\theta_2 \ll 1$, which seems to be an outcome of the configuration in which filament 2 ($\bm{n}_2$) is nearly parallel to filament 1 ($\bm{n}_1$). 
That is, collision between nearly parallel filaments leads to a massive clump irrespective of the direction of relative velocity.
\citet{2021MNRAS.507.3486H} (hydrodynamic models) 
and \citet{2023ApJ...954..129K} (MHD models) studied collisions between two parallel filaments with the relative velocity perpendicular to the filaments. 
Their setups belong to this configuration.
\item
Mass $F$ increases with decreasing $\phi_2 \ll 1$.
When $\phi_2\ll 1$, the three vectors, $\bm{n}_1$, $\bm{n}_2$, and $\bm{n}_v$, are approximately coplanar to the $x$-$z$ plane. 
At the beginning of collision, vectors $\bm{n}_1$ and $\bm{n}_2$ are not in a skew position.
If vector $\bm{n}_v$ is on the plane formed by the two vectors, where these three vectors are coplanar, 
the collision proceeds like closing a zipper (see figure 3 of \authorcite{10.1093/mnrasl/slae045} \yearcite{10.1093/mnrasl/slae045} for an example of this mode).
\item
Mass $F$ is large for $\theta_v \ll 1$.
In this case, filament 1 is extending nearly toward the relative velocity, $\bm{n}_v$.  
This induces a collision in which a large portion of filament 1 flows into filament 2 and a massive gas clump is expected.
The bulk motion of the filament appears to have the effect of gradually increasing the mass of the initially impinging gas, similar to the coplanar case discussed above.    
\end{enumerate}

\subsection{Global mass functions of filaments and hubs}
Figure \ref{fig.5} shows the expected hub mass distribution resulting from the collision of the filaments with a given line-mass $\lambda_0$. 
Here, we consider a global hub mass function, when the line-mass has a distribution in interstellar space. 

Figure \ref{fig.5} shows that the hub mass functions (MFs) are well fitted by the power-law.
The orthogonal filament-filament collisions form the minimum mass hubs, $M_\mathrm{hub}=4q\lambda_0$, from the filaments with the line-mass of $\lambda_0$.
If we fix the axis ratio $p/q$, the hub mass function (figure \ref{fig.5}) depends on $M_\mathrm{hub}/(4q\lambda_0)$.
The mass function is written as 
\begin{equation}
\frac{dN_M}{d\log M}=A\left(\frac{M}{4q\lambda_0}\right)^{+\gamma_M}, \label{eq:dNM/dlogM}
\end{equation}
where $\gamma_M=-3.78$ ($\Sigma=S$) or $-2.96$ ($\Sigma=S+T$). 
 
The line-mass function of filamentary structures is studied observationaly for less-massive and massive star-forming regions by \citet{2019A&A...629L...4A} and \citet{2025A&A...703A..74K
}, respectively [see also \citet{2021ApJ...916...83A}].
\citet{2019A&A...629L...4A} obtained the line-mass function of the filamentary structures found in the Gould's belt molecular clouds
\begin{equation}
\frac{dN_\lambda}{d\log\lambda_0}=B \lambda_0^{+\gamma_\lambda},\label{eq:dNlambda/dloglambda}
\end{equation}
for thermally supercritical filaments,
 $16 M_\odot\mathrm{pc}^{-1}\lesssim \lambda_0\lesssim 200 M_\odot\mathrm{pc}^{-1}$.
The power-law exponent is equal to $\gamma_\lambda=-1.5\pm0.1$.  
On the other hand, \citet{2025A&A...703A..74K
} measured three typical hub-filament systems found in massive star-forming regions and obtained the exponent as
 $\gamma_\lambda=-1.36\pm0.26$ [W3(OH)], $-1.04\pm 0.20$ (W3-Main), $-0.86\pm 0.17$ (S 106) for filaments outside the hubs,
 where the line-mass is smaller than $\lesssim 200M_\odot\mathrm{pc}^{-1}$.
However, they claimed that the power-law index in the hub regions are $\gamma_\lambda=-1.20\pm0.23$ [W3(OH)], $-2.21\pm 0.50$ (W3-Main), $-1.41\pm 0.28$ (S 106), where the typical line-mass is as large as $50 M_\odot \lesssim \lambda_0 \lesssim 1000 M_\odot\mathrm{pc}^{-1}$.

Considering collisions between filaments of equal line-mass, 
from equations (\ref{eq:dNM/dlogM}) and (\ref{eq:dNlambda/dloglambda}),
 the mass function contributed by the collision between the filaments of the line-mass $\lambda_0$ is  
\begin{align}
\frac{dN_\lambda}{d\log\lambda_0}\times \left(\frac{M}{4q\lambda_0}\right)^{+\gamma_M}
&\propto\lambda_0^{+\gamma_\lambda-\gamma_M}M^{+\gamma_M}\ \ \ (M\ge 4q\lambda_0),
\end{align}
where we assumed that $p$ and $q$ remain constant even as $\lambda_0$ changes.
Integrating this from $\lambda_0=M/4p$ to $\lambda_0=M/4q$,
\begin{equation}
\int_{\lambda_0=M/4p}^{\lambda_0=M/4q} B\lambda_0^{\gamma_\lambda}\left(\frac{M}{4q\lambda_0}\right)^{+\gamma_M} d\log \lambda_0\propto \left(\frac{M}{4q}\right)^{+\gamma_\lambda},
\end{equation}
where we used $(M/4q)^{+\gamma_\lambda-\gamma_M}\gg (M/4p)^{+\gamma_\lambda-\gamma_M}$.
We find that the hub mass function is proportional to $M^{+\gamma_\lambda}$, 
 if $\gamma_M<\gamma_\lambda<0$.
This means that the filament line-mass function and the hub mass function have the same power-law exponent $\gamma_\lambda$.
Therefore, we expect the mass function of the filament hubs averaged over different regions to be proportional to $\propto M^{-1.5}$, if $\gamma_\lambda\simeq -1.5$.

\section{Summary}\label{sec:5}
\begin{enumerate}
\item
We derived the collision cross section (CCS) for two identical cylindrical filaments of length $2p$, radius $q$, and line-mass $\lambda_0$.
In contrast to spherical clouds, the CCS for a filament depends on the orientation of the filament's axis and its relative velocity.
The CCS is composed of two parts: one whose cross-sectional area is proportional to $p^2$ [equation\,(\ref{eq:S})],
 and one whose cross-sectional area is proportional to $pq$\,[equation (\ref{eq:CCS})].  
If the distributions of the filament's axis and relative velocity are all isotropic,
 the CCS of a thin filament ($q \ll p$) is 1/8 of the CCS of a spherical cloud with radius $p$.    
\item
Based on the idea that molecular filaments arise from a disk structure,
 we studied several models in which the direction of the filament axis and that of the relative velocity are
 restricted rather than being isotropically distributed.  
We found that the closer all directions are to each other, the more pronounced the suppression of CCS becomes.
\item
We have derived a formula that approximates the hub mass left behind as a result of the filament collisions.
When two filaments collide, we model that the overlapping portion of the two filaments forms a hub.
The overlapping mass is expressed as equations (\ref{eq:Mhub}) and (\ref{eq:DM}).
A relatively massive hub forms either (1) when a collision occurs between two parallel filaments,
(2) when the axes of the filaments and their relative velocity vectors are coplanar,
or (3) when one of the filaments has the axis parallel to the relative velocity.   
\item
Using the CCS and the approximated hub mass, we calculated the hub mass function (figure \ref{fig.5}). 
The hub mass function fits well with a power-law function.
The power index is obtained as $\gamma_M\simeq -3.78$ for $\Sigma=S$ and $\gamma_M\simeq -2.96$ for  $\Sigma=S+T$,
 for $q/p=0.1$.
\item
The line-mass function of the filaments and the global mass function of the hubs have the same power-law exponent $\gamma_\lambda\simeq -1.5$.  
This means that a massive hub is formed by a collision between filaments with a large line-mass.  
\end{enumerate}

 \begin{ack}
Software: 
numpy (https://numpy.org/doc/stable/), matplotlib (https://matplotlib.org/),
and GNU Fortran (https://gcc.gnu.org/fortran/)
\end{ack}

\section*{Funding}
 This work was supported by JSPS KAKENHI Grant Numbers JP19K03919 (KT) and JP22J11106, JP25K17443 (RK).


\appendix 
\section{Allowable region of the impact parameter for two identical filaments with a finite width to collide}\label{sec:A1}

\begin{figure}[hbtp]
\begin{center}
\includegraphics[width=8cm]{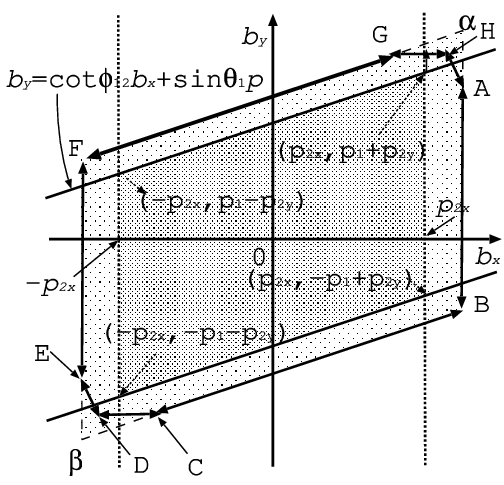}
\end{center}
\caption{Allowable region of the impact parameter $\bm{b}={}^t(b_x,\,b_y)$ when two filaments with a finite width collide each other.
The inner shaded parallelogram corresponds to the model with no width ($q \ll p$; section \ref{ssec:2.1}).
The outer octagon represents the additional CCS for the filaments with a finite width ($q \lesssim p$; section \ref{ssec:2.2}).
{Alt text: Two co-centered parallelograms.
Inner parallelogram is shaded deeply, while the peripheral is lightly shaded.  
The outer parallelogram closely approximates the octagon representing the exact collision cross section. }
}
\label{fig.A1}
\end{figure}
In figure \ref{fig.2}, two filaments are projected onto a plane perpendicular to the relative velocity. 
In the figure, the four vertices of each filament are labeled UL, UR, LR, and LL.
Filament 1 is aligned along the $y$-axis.
The coordinates of the four vertices of filament 1 are expressed as
\begin{align}
&\mathrm{UR\ }\begin{pmatrix}+q\\+p'_1\end{pmatrix},\ 
\mathrm{UL\ }\begin{pmatrix}-q\\+p'_1\end{pmatrix}, \ 
\mathrm{LR\ }\begin{pmatrix}+q\\-p'_1\end{pmatrix},\ 
\mathrm{LL\ }\begin{pmatrix}-q\\-p'_1\end{pmatrix},
\end{align}
where $p'_1=|\bm{p}'_1|=p|\bm{n}'_1|$.
The same description applies to filament 2 as 
\begin{align}
&\mathrm{UR\ }\begin{pmatrix}b_x+p'_{2x}+\cos\phi_{12}\,q\\ b_y+p'_{2y}-\sin\phi_{12}\,q\end{pmatrix},
\mathrm{UL\ }\begin{pmatrix}b_x+p'_{2x}-\cos\phi_{12}\,q\\ b_y+p'_{2y}+\sin\phi_{12}\,q\end{pmatrix},\nonumber\\
&\mathrm{LR\ }\begin{pmatrix}b_x-p'_{2x}+\cos\phi_{12}\,q\\ b_y-p'_{2y}-\sin\phi_{12}\,q\end{pmatrix},
\mathrm{LL\ }\begin{pmatrix}b_x-p'_{2x}-\cos\phi_{12}\,q\\ b_y-p'_{2y}+\sin\phi_{12}\,q\end{pmatrix},
\end{align}
where $^t(p'_{2x}, p'_{2y})=\bm{p}'_2=p\bm{n}'_2=p'_2{}^t(\sin\phi_{12},\ \cos\phi_{12})$
and $\phi_{12}$ represents the angle between $\bm{n}'_1$ and $\bm{n}'_2$.

To determine the conditions under which two filaments overlap,
 we find the conditions under which the vertex of one filament overlaps the edge of the other filament.  

As an example, we consider vertex LL of filament 2 and the right edge of filament 1 (line segment UR-LR).
The conditions are as follows:
\begin{equation}
\left\{\begin{aligned}
&b_x-p'_{2x}-\cos\phi_{12}\,q=q,\\
&-p'_1\le b_y-p'_{2y}+\sin\phi_{12}\,q \le +p'_1.
\end{aligned}
\right.
\end{equation}
Thus,
\begin{equation}
\left\{\begin{aligned}
&b_x=p'_{2x}+(1+\cos\phi_{12})q,\\
-p'_1 + p'_{2y}-\sin\phi_{12}\,q \le &b_y \le +p'_1+p'_{2y}-\sin\phi_{12}\,q.
\end{aligned}
\right.
\label{eq:part-iii}
\end{equation}
This indicates the rightmost vertical line segment AB in figure \ref{fig.A1},
 of which the end points are $^t(b_x,b_y)={}^t[p'_{2x}+(1+\cos\phi_{12})q,\ p'_1+p'_{2y}-\sin\phi_{12}\,q]$
 and $^t[p'_{2x}+(1+\cos\phi_{12})q,\ -p'_1+p'_{2y}-\sin\phi_{12}\,q]$.

Next, we consider the combination of vertex LR of filament 1 and the line segment LL-UL on filament 2.
The coordinate of the new vertex C is $^t[-p'_{2x}+(1+\cos\phi_{12})q,\ -p'_1-p'_{2y}-\sin\phi_{12}\,q]$.
We continue to calculate the coordinates of the vertices in a similar manner, which are shown in table \ref{tbl.A1}.
The coordinates of other vertices not shown in the table can be easily determined because the figure
 is point-symmetric with respect to the origin.

\begin{table}[htp]
\caption{Coordinate of vertices.}
{\centering\tabcolsep = 2pt
\begin{tabular}{llll}
P$^*$\hspace*{-2mm} & F1$^\dag$\hspace*{-2mm} & F2$^\ddag$\hspace*{-2mm} & Coordinate in figure \ref{fig.2}\\
\hline
\hline
A & UR & LL & $^t[p'_{2x}+(1+\cos\phi_{12})q,\ p'_1+p'_{2y}-\sin\phi_{12}\,q]$ \\
B & LR & LL & $^t[p'_{2x}+(1+\cos\phi_{12})q,\ -p'_1+p'_{2y}-\sin\phi_{12}\,q]$ \\
C & LR & UL & $^t[-p'_{2x}+(1+\cos\phi_{12})q,\ -p'_1-p'_{2y}-\sin\phi_{12}\,q]$ \\
D & LL & UL & $^t[-p'_{2x}-(1-\cos\phi_{12})q,\ -p'_1- p'_{2y}-\sin\phi_{12}\,q]$  \\ 
E$^\#$ & LL & UR & $^t[-p'_{2x}-(1+\cos\phi_{12})q,\ -p'_1-p'_{2y}+\sin\phi_{12}\,q]$ \\
\hline
\end{tabular}
}
$^*$Vertex in figure \ref{fig.A1}.
$^\dag$Vertex of the filament 1 in figure \ref{fig.2}.
$^\ddag$Vertex of the filament 2 in figure \ref{fig.2}.
$^\#$Point-symmetric with point A with respect to the origin.
\label{tbl.A1}
\end{table}%

To derive the CCS, we calculated the area of the octagon ABCDEFGH in figure \ref{fig.A1}.
First, we calculate the area of the parallelogram $\alpha$B$\beta$F (within the lightly shaded area).     
The coordinate of point $\alpha$ is obtained from the intersection
 between the extension of line segment AB and that of line segment FG as follows:
\begin{align}
\bm{r}_\alpha
=\begin{pmatrix}
p'_{2x}+(1+\cos\phi_{12})q\\
p'_1+p'_{2y}+\sin\phi_{12}\,q+2\frac{p'_{2y}}{p'_{2x}}(1+\cos\phi_{12})q
\end{pmatrix}.
\end{align} 
The length of the base of the parallelogram (line segment $\alpha$B) is equal to 
 $2p'_1+2\sin\phi_{12}\,q+2(p'_{2y}/p'_{2x})(1+\cos\phi_{12})q$ and the height
 is $2p'_{2x}+2(1+\cos\phi_{12})q$.
Thus, the area is 
\begin{align}
A_{\alpha\mathrm{B}\beta\mathrm{F}}
=&4\sin\phi_{12}p'_1p'_2+4(1+\cos\phi_{12})(p'_1+p'_2)q\nonumber\\
&+4[(1+\cos\phi_{12})^2/\sin\phi_{12}]q^2.
\end{align}

Next, the areas of triangles $\alpha$GH and $\alpha$HA are 
\begin{align}
\triangle{\alpha\mathrm{GH}}=
\triangle{\alpha\mathrm{HA}}
&=2(1+\cos\phi_{12})\cot\phi_{12}\,q^2.
\end{align}
The area of the octagon is given as
\begin{align}
A_\mathrm{oct}=&A_{\alpha\mathrm{B}\beta\mathrm{F}}-2(\triangle{\alpha\mathrm{GH}}+\triangle{\alpha\mathrm{HA}})\nonumber\\
=&4\sin\phi_{12}p'_1p'_2+4(1+\cos\phi_{12})(p'_1+p'_2)q
+4\cot\phi_{12}\,q^2.\label{eq:A1CCS}
\end{align}
The first term represents the darkly shaded area in figure \ref{fig.A1} and corresponds to the CCS
 of a widthless filament.
The second term represents approximately the lightly shaded area and arises from the finite width of the filament. 
Neglecting the term $\propto q^2$, the CCS between two identical filaments with a finite width is given by equation (\ref{eq:CCS}).

\section{Overlap mass for crossing two cylinders}\label{sec:A2}

\begin{figure}[hbtp]
\begin{center}
\includegraphics[width=80mm]{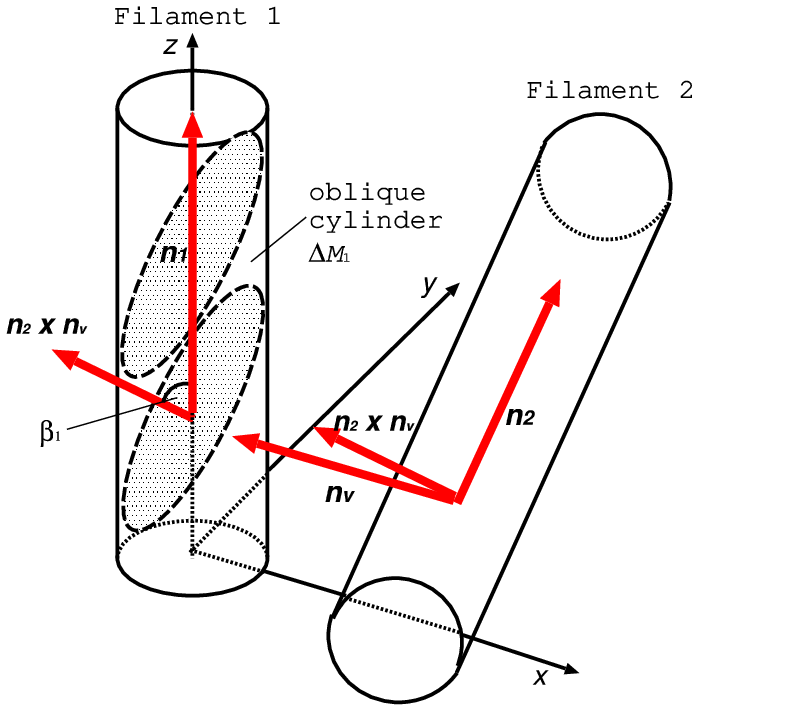}
\end{center}
\caption{Oblique collision. 
The two filaments are neither parallel nor perpendicular to each other.
Vectors $\bm{n}_1$,  $\bm{n}_2$, and $\bm{n}_v$ represent the axes of the two filaments and the direction of their relative velocity, respectively.
The mass of  $\Delta M_1$ of the portion of filament 1 swept by filament 2 is obtained
as the mass of an oblique cylinder with the top and bottom surfaces shaded in the figure. 
Angle $\beta_1$ is the angle between vector $\bm{n}_{c2}=(\bm{n}_v\times\bm{n}_2)/|\bm{n}_v\times\bm{n}_2|$ and  
vector $\bm{n}_1$.
{Alt text: A schematic diagram which explains the overlapping portion when two regular cylinders are passing.}
}
\label{fig.A2}
\end{figure}

First, we derive the mass of $\Delta M_1$ of the portion of filament 1 swept by filament 2.
The shaded ellipse in figure \ref{fig.A2} is a plane (cutting plane) determined
 by the axis $\bm{n}_2$ of filament 2 and its direction of movement $\bm{n}_v$.
We calculate the volume of the portion (oblique cylinder) cut out from filament 1 by the two cutting planes
 separated by a perpendicular distance $2q$. 
The normal vector perpendicular to this cutting plane is written as
\begin{equation}
\bm{n}_{c2}\equiv\frac{\bm{n}_v\times \bm{n}_2}{\left|\bm{n}_v\times \bm{n}_2\right|}.
\end{equation}

The angle $\beta_1$ between the normal vector, $\bm{n}_{c2}$, and the axis of filament 1, $\bm{n}_1$,
 controls the volume of the overlapping region of filament 1,
 which is the shaded oblique cylinder. 
That is, if $\beta_1=0$, in other words, if filament 1 is cut perpendicular to the axis,
 the overlapping portion is a right circular cylinder with height $2q$. 
If $0 < \beta_1< \pi/2$, the overlapping region is an oblique cylinder with 
 slant height $\ell_1 =2q/\cos\beta_1$.
Therefore, the volume and mass of the oblique cylinder are
 expressed as
\begin{align}
 &\Delta V_1=\ell_1 \pi q^2=2\pi q^3/\cos\beta_1,\\
 &\Delta M_1=\ell_1 \lambda_0=2q\lambda_0/\cos\beta_1.\label{eq:DM1}
\end{align} 

By repeating the same calculation for the mass of filament 2 swept by filament 1,
the volume and mass are as follows:  
\begin{align}
 &\Delta V_2=\ell_2 \pi q^2=2\pi q^3/\cos\beta_2,\\
 &\Delta M_2=\ell_2 \lambda_0=2q\lambda_0/\cos\beta_2, \label{eq:DM2}
\end{align} 
where $\ell_2$ and $\beta_2$ represent the slant height of the oblique cylinder
and the angle between $\bm{n}_2$ and $\bm{n}_{c1}=(\bm{n}_1\times\bm{n}_v)/|\bm{n}_1\times\bm{n}_v|$, respectively. 

From equations (\ref{eq:DM1}) and (\ref{eq:DM2}),
 the overlapping mass of two identical cylinders is obtained as
\begin{equation}
\Delta M=2q\lambda_0 \left(\frac{1}{\cos\beta_1}+\frac{1}{\cos\beta_2}\right).\label{eq:DM}
\end{equation}
 



\bibliography{tomisaka_v2}{}
\bibliographystyle{aasjournal}
\end{document}